\documentclass[%
 aip,
 jmp,%
 amsmath,amssymb,
 reprint,%
]{revtex4-2}

\usepackage{graphicx}% Include figure files
\usepackage{dcolumn}% Align table columns on decimal point
\usepackage{bm}% bold math
\preprint{ }
 \usepackage[version=4]{mhchem}
\usepackage[normalem]{ulem}
\usepackage{graphicx}% Include figure files
\usepackage{dcolumn,xcolor}% Align table columns on decimal point
\usepackage{bm}% bold math
\begin{document}

\preprint{}

\title[Q-plate readout in reflection]{Structure orientation determined in transmission and reflection: q-plate}% Force line breaks with \\
%\thanks{Footnote to title of article.}
\author{%\blue{Open List,} 
Hsin-Hui Huang$^{1,2,3*}$, Meguya Ryu$^{4*}$, Shuji Kamegaki$^4$, Haoran Mu$^{1,3}$, Eulalia Puig Vilardell$^5$, Vijayakumar Anand$^1$, Jitraporn Vongsvivut$^6$, Nguyen Hoai An Le$^2$, Tomas Katkus$^2$, 
Gediminas Seniutinas$^7$, Junko Morikawa$^{4,8,9}$, Saulius Juodkazis$^{1,5,8}$
}%

\affiliation{Optical Sciences Centre, 
Swinburne University of Technology, Hawthorn, Victoria 3122, Australia}
\affiliation{Australian Research Council (ARC) Industrial Transformation Training Centre in Surface Engineering for Advanced Materials (SEAM), Swinburne University of Technology, Hawthorn, VIC, 3122, Australia}
\affiliation{Melbourne Center for Nanofabrication (MCN), 151 Wellington Road, Clayton, Vic 3168, Australia}
\affiliation{~School of Materials and Chemical Technology, Institute of Science Tokyo, 2-12-1, Ookayama, Meguro-ku, Tokyo 152-8550, Japan}
\affiliation{Laser Research Center, Physics Faculty, Vilnius University, Saul\.{e}tekio Ave. 10, 10223 Vilnius, Lithuania}
\affiliation{Infrared Microspectroscopy (IRM) Beamline, ANSTO‒Australian Synchrotron, 800 Blackburn Road, Clayton, Victoria 3168, Australia}
\affiliation{Paul Scherrer Institut, Villigen PSI 5232, Switzerland
%\blue{Qnami AG, CH-4132, Muttenz, Switzerland}
}
\affiliation{~World Research Hub (WRH), School of Materials and Chemical Technology, Institute of Science Tokyo, 2-12-1, Ookayama, Meguro-ku, Tokyo 152-8550, Japan}
\affiliation{~Research Center for Autonomous Systems Materialogy (ASMat), Institute of Innovative Research, Institute of Science Tokyo,Yokohama 226-8501, Japan}

%\homepage{http://www.Second.institution.edu/~Charlie.Author.}
%\affiliation{%Second institution and/or address%\\This line break forced% with \\}%

\thanks{*Correspondence: H-H. H. hsinhuihuang@swin.edu.au; M. R. ryu.meguya@mct.isct.ac.jp  }

\date{\today}% It is always \today, today,

\begin{abstract}
Determination of orientation in the imaged sample/scene has a large application potential when the anisotropy of properties is analysed, usually, under a linearly polarised illumination. This study combined several improvements of microscopy imaging: use of an incoherent white illumination source (a lamp) with a spectral filter to define a spectral window, a plastic circular polariser to image with circularly polarised light (instead of linear), and a 4-pol. camera with integrated polarisers for simultaneous acquisition of images at four $\pi/4$-azimuth shifts. In the transmission mode, a high-fidelity readout of form birefringent optical elements, q-plates, was achieved using a fitting procedure based on the analytical expression of $S_0$ (intensity) Stokes parameters at the pixel level. In the reflection mode, the $S_0$ fit was used to determine the azimuth orientation of the q-plates, as well as %$2\theta_0$ of the q-plate segments was achieved using the same protocol as well as 
a generic $Amp\times\cos(2\vartheta-2\theta_0)+\mathrm{Offset}$ fit (at pixel level) applied to images taken at four $\vartheta$ azimuths. %Influence of light handedness and its change upon reflection as well as Stokes image processing as well as 
The 4-pol. analysis in reflection under circularly polarised illumination is discussed.  
\end{abstract}

%Use showkeys class option if keyword
\maketitle
\tableofcontents
\begin{quotation}
 %``lead paragraph'' 
\end{quotation}

\section{\label{intro}Introduction}

Detections of absorbance dichroism and birefringence are usually carried out using linearly polarised illumination at four (or more) polarisation azimuths (4+pol.) in transmission~\cite{Otani,Otani1}. The application potential of detecting anisotropies in absorbance and refractive index (real part) from reflection is much higher, including fields from space-based Earth observation~\cite{Ottaviani_2019,Chen_2025b} to microscopy for material science~\cite{Mehta_2013,Song_2021,Kamegaki_2024a} and medicine~\cite{Jacques_2016,He_2021a,Guan_2025}. Especially, if the resolution better than the diffraction limit could be achieved as shown in transmission~\cite{19n732,25nse70099}. With the currently available 4-pol. CCD cameras with pixel-level integrated polarisers, it was demonstrated that all four Stokes parameters~\cite{Schaefer}, hence, the state of polarisation, can be measured in one camera shot in transmission using circularly polarised illumination of the sample~\cite{24ap2300471}. This was demonstrated for microscopy and can be used for \emph{in situ} monitoring of amorphous to crystalline transformation in organic crystals~\cite{23lpr2200535}. Generation of circular/elliptical polarisation source and cross-polarised imaging is well established~\cite{Shribak}. Metasurfaces with polarisation encoded analysis were proposed for polarisation resolved imaging~\cite{Li23}.
Formation of ordering and recognition of anisotropic patterns can be achieved since several measurements to establish Stokes parameters are not required. In Earth's observation context by circularly polarised radio waves (or other EM radiation), analysis of reflected light for absorption, scattering, reflection anisotropies has high application potential. Single data acquisition is required due to the fast-moving image frame (LEO orbit is considered). Single-shot 3D imaging is another direction for the detection of fast events~\cite{Maeda}. 

In earlier studies of 4+pol. method in transmission mode, a circularly polarised light was used to illuminate onto the sample and a 4-pol. camera was used to detect the transmitted power. The analytical solution for the normalised Stokes parameter $S_0$ (full power) for the case when there is no dichroism or optical activity (birefringence for circularly polarised light)~\cite{24ap2300471}:
\begin{equation}\label{e1}
  S_0^{(out)}(\phi,\delta,\theta_0) = \frac{1}{2} - \frac{1}{2}\sin\delta\sin{[2(\phi - \theta_0)]},
\end{equation}
\noindent where $\phi$ is the azimuth angle, $\delta = 2\pi\Delta n H/\lambda$ is the retardance, and $\theta_0$ is the angle of maximum $T$; $\Delta n = n_e - n_o$ is the birefringence defined by the extraordinary and ordinary refractive indices $n_{e,o}$, respectively, $H$ is the length of birefringent region, $\lambda$ is the wavelength of light. By fitting the measured intensity by $Amp\times S_0$ (Eqn.~\ref{e1}) at four angles (pixels with on-chip polarisers), the amplitude $Amp$, retardance $\delta$, and its azimuth $\theta_0$ can be determined (three fitting parameters for four independent polarisation measurements with 4-pol. camera). 

The conjecture to rely on the only $S_0$ Stokes parameter (intensity), makes a change of handedness of circular polarisation upon reflection not critical and simplifies the analysis. Therefore, such 4+pol. methodology can be applied in the reflectance setup using circularly polarised illumination. This is the aim of the current study, with further simplifications by use of simple plastic circular polarisers and spectral filters to select a wavelength from the microscope's condenser illumination. %\red{/} 
The test sample for orientation determination by 4+pol. method in this study was a q-plate~\cite{Marrucci,16aom306}, the form-birefringent azimuthally changing nano-grating structure defined by the electron beam lithography (EBL) on/in a 5-$\mu$m-thick poly-crystalline diamond membrane~\cite{18mt798}. Such q-plate acts as an optical spin-orbit converter with the optical slow-axis rotating with azimuth angle $\theta$ according to $\beta = q\times\theta$. This optical element generates an optical vortex beam with topological charge $\phi = \pm 2q$, i.e., the number of rotations along propagation of a single wavelength. In imaging, such vortex beams leverage helical wavefronts and phase singularities to improve contrast and depth sensitivity beyond the limits of conventional Gaussian illumination~\cite{1v,2V,3v,9v,10v,11v}. These structured beams interact with sample morphology to encode volumetric features and optical anisotropies, which are essential for high-resolution 3D reconstruction~\cite{12v,13v}. Accordingly, the form‑birefringent q‑plate used here serves not only as a well‑defined test structure for polarization‑based orientation determination but also as a representative generator of OAM beams relevant to modern multidimensional imaging.

Devices such as q‑plates consisting of patterned anisotropic optical elements with azimuthally varying optical‑axis orientation are widely used to produce vortex beams with controlled topological charge and polarization states, enabling both scalar and vector structured beams with tailored phase and polarization topology. Conventional methods of generation of orbital angular momentum (OAM) beams involve grayscale phase distributions with spiral phase plates (SPPs), which are challenging as they require continuous thickness variation and often have less modal purity~\cite{4v}, and alternative methods involving computer‑generated holograms such as fork gratings have improved modal purity but low diffraction efficiency~\cite{5v}. Form birefringence, on the other hand, is based on anisotropic micro‑ or nano‑structured materials that impart space‑variant phase shifts converting the spin angular momentum of circularly polarized light into OAM via a geometric‑phase mechanism~\cite{6v,7v,8v}, making it a key method for generating structured light beams carrying OAM.

The most efficient spin-orbit conversion upon illumination with circularly polarised light occurs at a longitudinal phase retardance of $\pi$ between E-field components along the grating pattern $\parallel$ (extraordinary $e$-beam) and perpendicular $\perp$ to it (ordinary $o$-beam). For such condition, $\delta = 2\pi(n_e - n_0)H/\lambda$ for the height $H$ of the structure. The form-birefringence $\Delta n$ is defining the phase retardance and is dependent on the refractive index of the host material (diamond $n = 2.45$ at 425~nm) and the fill-factor of the grating pattern.

%\red{Simple 4+pol. setup based on filters, plastic circular polarisers and 4-pol. camera was used for transmittance and reflectance analysis of anisotropy (in this case form-birefringence).}
This study reports on simplification of polarisation analysis using optical microscopy at visible spectral range by implementing simple plastic polarisers, spectral bandpass filters over condenser illumination, and 4-pol. camera with on-chip integrated linear polarisers at 45$^\circ$ azimuthal differences between them. In transmittance mode, reliable determination of orientation of q-plate form birefringent regions was achieved. For reflection mode of operation, $S_0$ (Eqn.~\ref{e1}) analysis of structure orientation % helicity change upon reflectance has to be accounted for 
showed \emph{quantitative} agreement, which had an uncertainty of $\pi$-phase change at azimuthal angles close to $\pm\pi$. %however, 
Also, the orientation of the form-birefringent segments was retrieved by a simple fit to $\cos(2\vartheta)$ for azimuthal $\vartheta$ dependence. 

\section{Samples and Methods}

A standard up-right microscope (Olympus BX51, Olympus, Ishikawa, Japan) was used for experiment with 4-pol. camera (CMOS CS505MUP1, Thorlabs Inc., Newton, NJ, USA) and plastic circular polarisers for the right/left hand circular (RHC/LHC) spectrally broad band 400-700~nm wavelength operation (Circular Polarizers (CP42HER and CP42HE, Edmund Optics, Barrington, NJ, USA; $\sim 40\%$ transmittance ). For measurements, spectrally narrow $\sim 10$~nm band pass filter (Nikon) at central wavelength 425~nm was used by placing it directly on the condenser illuminator port for transmission or reflection modes of measurement as used previously~\cite{18sr17652}. 

%____________________________fig 1
\begin{figure*}[b!]
\centering\includegraphics[width=1\textwidth]{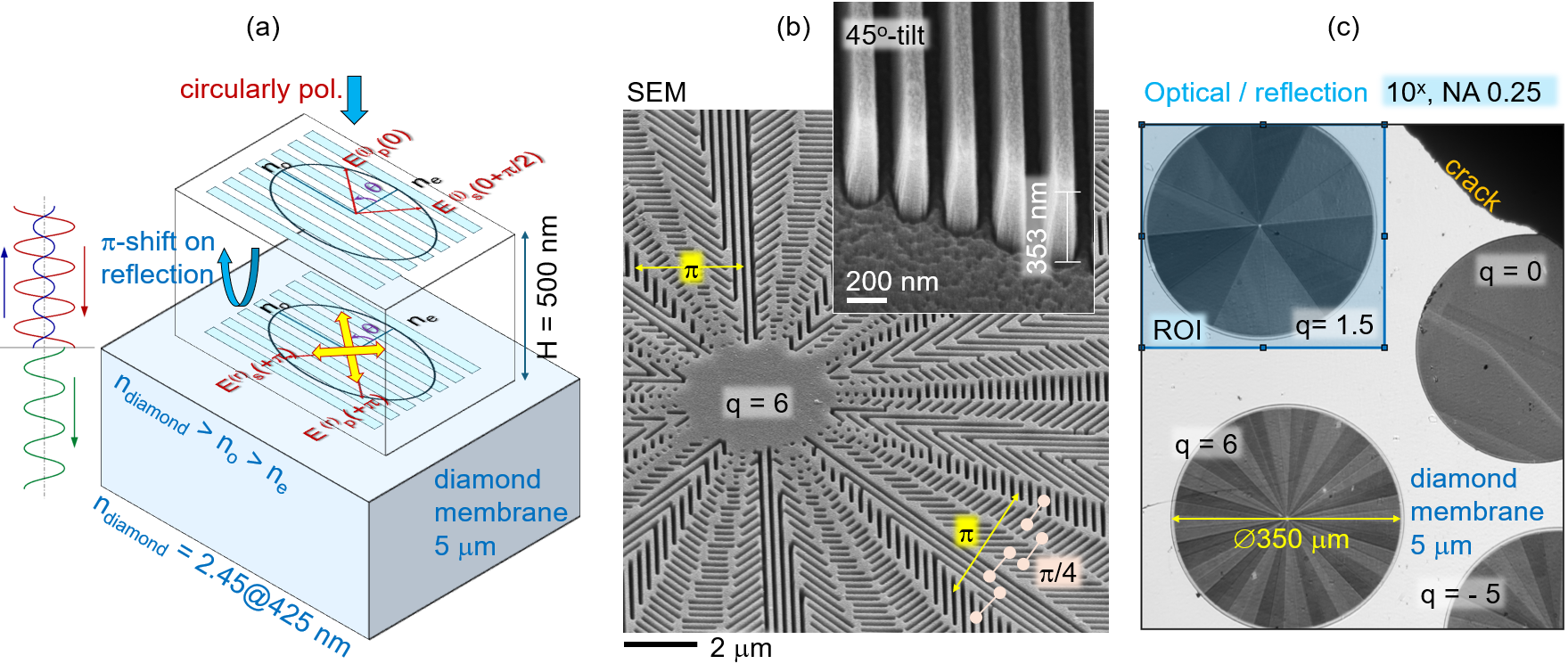}
\caption{\label{f-sem} Q-plates: azimuthally changing birefringence. The slow-axis follows azimuth angle $\varphi$ as $\beta = q\varphi$. (a) Schematics of a form-birefringent structure made in diamond micro-membrane and illuminated with circularly polarised light; parameters used in this study are shown. The refractive index of extraordinary beam is smaller than that of ordinary $n_e < n_o$. (b) SEM image of an azimuthally changing form-birefringent structure of a q-plate $q=6$ (generator of optical vortex with topological charge $2q$). (c) Optical image of few different q-plates on the membrane. }
\end{figure*}

The 4-pol. camera available for visible spectral range improves polarisation analysis, makes it real time since there is no need to acquire separate images under polarised light illumination for Stokes polarimetry or spectroscopy. Image vignetting when wire grid polarisers are used in IR for 4+pol. analysis makes it computationally intensive~\cite{21as1544}. This was not required for current study and images were well matched on the pixel level. This is critically important when a pixel level fitting was used for orientation and birefringence determination in this study. 

%____________________________fig 2
\begin{figure*}[t!]
\centering\includegraphics[width=1\textwidth]{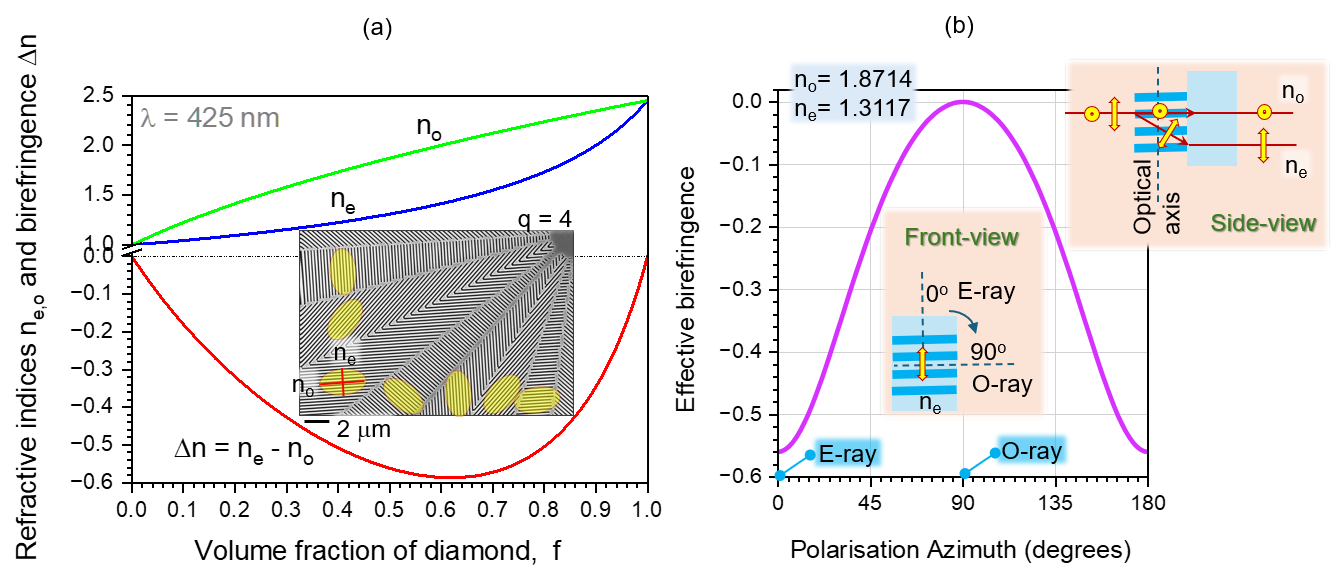}
\caption{\label{f-fb} (a) Form birefringence of nano-gratings in diamond at different filing ratio $f$: $\Delta n = n_e - n_o < 0$; diamond's $n = 2.45$ and $\lambda = 425$~nm. Inset shows SEM image of $q=4$ plate with shematics of refractive index ellipsis (slow-axis is along $n_o$). Period of structure $\Lambda = 250$~nm. (b) Effective form-birefringence for the normal incidence at different azimuth ($0^\circ$ corresponds to the optical axis). Calculated by Eq.~\ref{e-az} for the volume fraction $f = 0.5$ and $n_e = 1.8714$ and $n_e = 1.3117$ shown in (a). }
\end{figure*}

Image analysis was carried out using own-developed Matlab code to make fitting at the pixel level by Eq.~\ref{e1} from the acquired 4-pol. images. Program was running on a desktop computer in parallelized version of the program since all fitting is independent for each pixel. For a VGA $640\times 480$ pixels data set, the parallel mode of calculation shortened the data processing time by $\sim 20$ times as compared with sequential mode. %\red{[....relevant details to add.....] }

Q-plate samples were defined by electron-beam lithography (EBL) and plasma etched on free-standing polycrystalline diamond membranes~\cite{18mt798}. High quality diamond 2D membranes as well as 3D millimetres-sized elements grown by chemical vapour deposition (CVD) have become widely available and used in quantum optics applications~\cite{Jiang}. In all measurements $T-$ and $R-$modes, the same sample orientation facing up-wards with q-plate to the 4-pol. camera was maintained. 

\section{Results}

Figure~\ref{f-sem} shows the structure of form birefringent pattern of q-plates made on $5~\mu$m-thick diamond membrane imaged by electron and optical microscopies. When structure is illuminated from the top in reflectance measurements, there are $\pi$ phase changes upon reflection when light propagates from low-to-high refractive index. Since the refractive index of diamond $n_d\simeq 2.45$ at $\lambda = 425$~nm used in this study and $n_d>n_o>n_e$, there is $\pi$-phase shift also at the diamond substrate and form-birefringent q-plate at its bottom (Fig.~\ref{f-sem}(a)).

%____________________________fig 3
\begin{figure*}[t!]
\centering\includegraphics[width=1\textwidth]{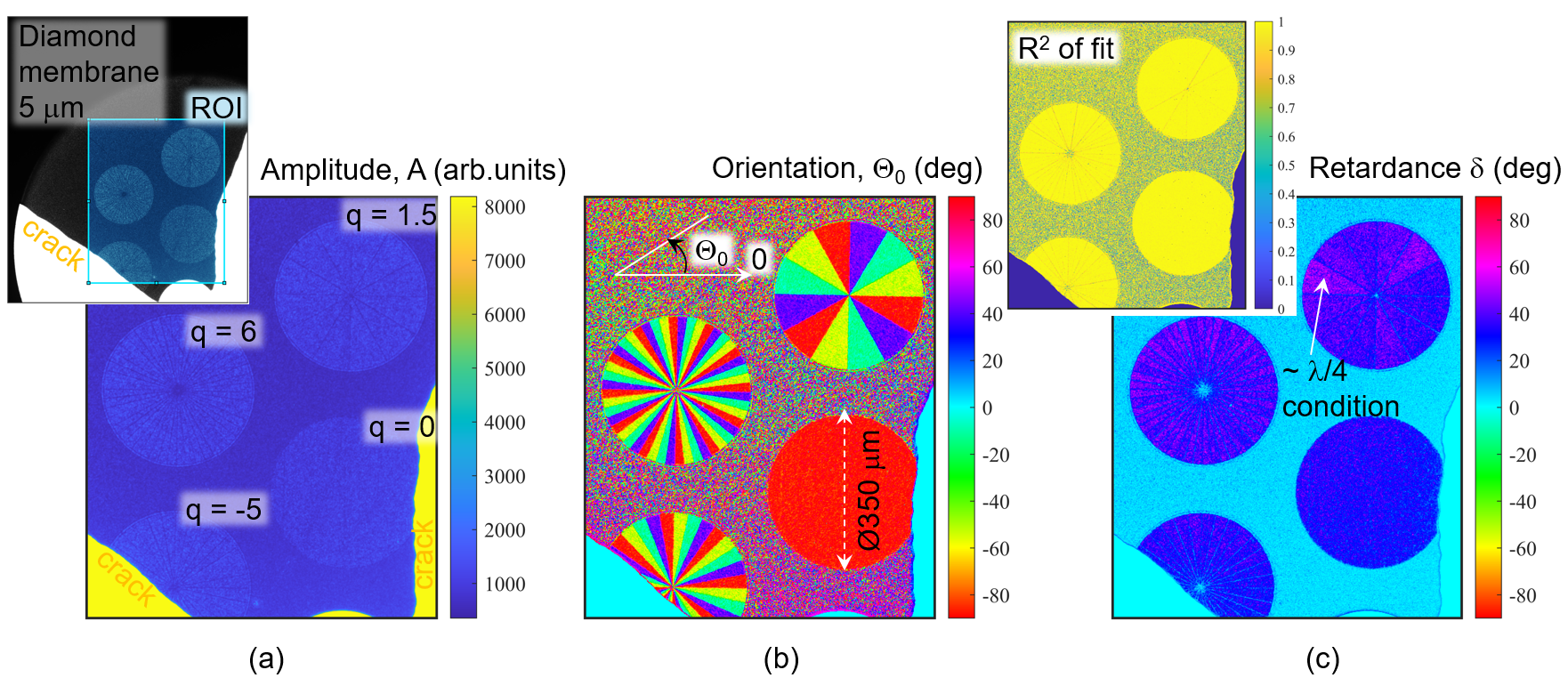}
\caption{\label{f-T} Transmittance $T$ of q-plates imaged by 4-pol. camera at numerical aperture $5^\times NA = 0.13$
and fitted by $Fit(\varphi) = Amp[(1 - \sin\delta\sin[2(\varphi - \theta_0)])/2]$ (see panels (a-c) for the fit parameters
amplitude, orientation (highest T), and retardance, respectively: $Amp, \theta_0, \delta$ calculated for each pixel). %Condenser lamp was covered by a 425 nm band pass filter ($\sim$20~nm FWHM) and RHC polarisation was set by a single plastic element (Edmund) consisting of a linear polariser sandwiched with a quarter waveplate (broad band 400 - 700 nm). Sample was placed right after the polariser on a sample stage of microscope facing up-wards with q-plate to 4-pol. camera.
}
\end{figure*}

The form birefringence depends to the volume fraction of material $f$ and can be calculated from the known isotropic refractive index of diamond for light's E-field oscillating along the extraordinary (along optical axis) and ordinary directions~\cite{14pqe119}:
\begin{equation}\label{Eq1}
n_e=\sqrt{\frac{n_1^2n_2^2}{(1-f)n_2^2+fn_1^2}}; ~~~~ n_o=\sqrt{(1-f)n_1^2+fn_2^2},  
\end{equation}
\noindent where $n_{1,2}$ are the refractive indices of environment (air) and diamond, respectively. Figure~\ref{f-fb} shows Eqn.~\ref{Eq1} for 425~nm used in this study.

For normal incidence onto a q-plate, different segments are probed at different azimuth $\vartheta$ in respect to the fast-axis when linearly polarised light is illuminated (an extraordinary beam with E-field oscillation along the optical axis; $n_e$). The effective birefringence is given by expression adopted from uniaxial birefringence~\cite{Veiras}:
\begin{equation}\label{e-az}
 \Delta n^{'}(\vartheta) = n_e\times \left[1-\frac{n_o}{\sqrt{n_o^2\sin^2\vartheta + n_e^2\cos^2\vartheta}} \right]. 
\end{equation}
Figure~\ref{f-fb}(b) shows the effective birefringence $\Delta n^{'}$ for the form birefringent structure with $f=0.5$, which corresponds to the $n_e = 1.3117$ and $n_o = 1.8714$. The largest birefringent is for $\vartheta = 0^\circ$ (E-ray) and $\Delta n^{'}\rightarrow 0$ for the ordinary beam (O-ray) at $\vartheta = 90^\circ$.

\subsection{Orientation determined from the fit of transmitted power}

Possibility to determine orientation of q-plate segments using a fit by $Amp\times S_0$ (Eq.~\ref{e1}) of the transmitted light is straight forward from 4-pol. camera images at $\phi = 0^\circ, 45^\circ, 90^\circ, 135^\circ$ as shown in Fig.~\ref{f-T}. The fit has high fidelity with $R^2>0.9$. Orientation of q-segments which are by $\pi/4$ orientation rotated between the neighbouring segments is well retrieved. The retardance on the q-plates reads as $\delta \simeq 30^\circ$. Considering form birefringence of $\Delta n\simeq 0.5$ (large duty cycle close to $0.5$ of the pattern; Fig.~\ref{f-fb}), $H = 500$~nm, $\lambda=425$~nm: $\delta = 211.8^\circ$ or $31.8^\circ$ ($\pi$-folded), which is closely matching the experimentally determined $\delta$ from the fit (Fig.~\ref{f-T}(c)). As expected, the diamond membrane regions without q-plate as well as empty (no sample) locations showed $\delta = 0$. The region of q-plate in one colour (red) is for the $q = 0$ plate (a grating) with orientation $\theta_0=\pm 90^\circ$ (Fig.~\ref{f-T}(b)); see Fig.~\ref{f-q0} for larger resolution and magnification. The amplitude $Amp$ is uniform across all the area of q-plate with many segments, which shows high fidelity of the fabricated pattern.

The q-plate had a comparatively small retardance, which makes spin-orbital conversion of the incoming RHC $s=+1$ into LHC $s = -1$ with a compensating orbital momentum $l=+2$ not efficient. For the most efficient full spin-to-orbital conversion the q-plate has to fulfill the $\lambda/2$-plate ($\pi$) condition. The uniform regions for all fit parameters over the q-plates' regions confirms a homogeneous performance corresponding to the dominant polarisation. 

%____________________________fig 4
\begin{figure*}[tb]
\centering\includegraphics[width=1\textwidth]{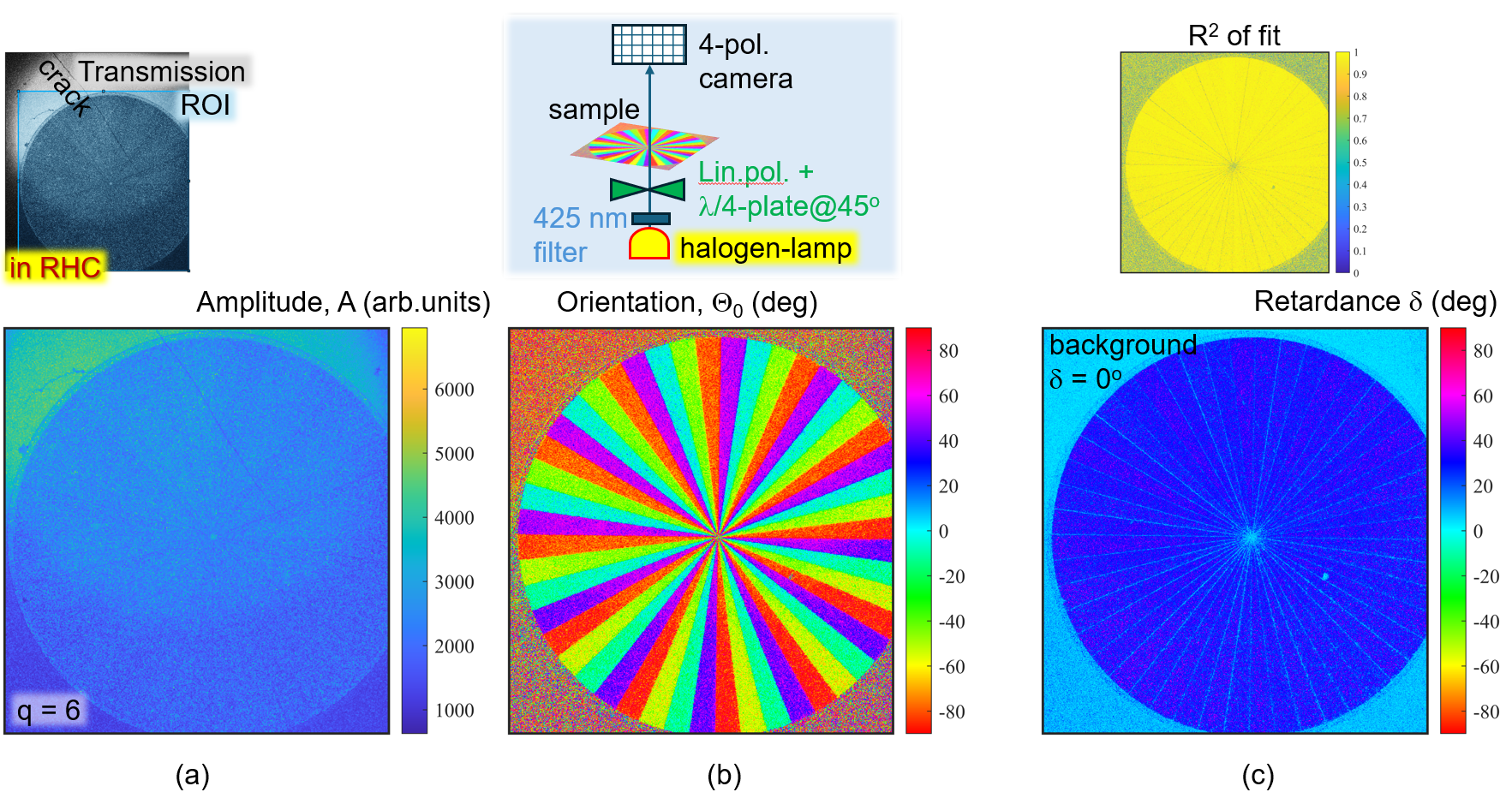}
\caption{\label{f-qu6T} Transmittance $T$ of $q = 6$ plates imaged by 4-pol. camera at numerical aperture $20^\times NA = 0.5$ and fitted by $Fit(\varphi)=Amp[(1-\sin\delta\sin{[2(\varphi - \theta_0)]})/2]$ (see panels (a-c) for the fit parameters amplitude, orientation (highest $T$), and retardance, respectively: $Amp, \theta_0, \delta$ calculated for each pixel). Top-inset in (b) shows geometry of measurements with sample's q-plate facing 4-pol. camera (up-wards position). Condenser lamp was covered by a 425~nm band pass filter ($\sim 20$~nm FWHM) and RHC polarisation was set by a single plastic element consisting of a linear polariser sandwiched with a quarter waveplate (broad band 400-700 nm). Sample was placed right after the polariser on a sample stage of microscope.
}
\end{figure*}

\subsection{Orientation from fit of reflected power}

\begin{figure*}[tb]
\centering\includegraphics[width=1\textwidth]{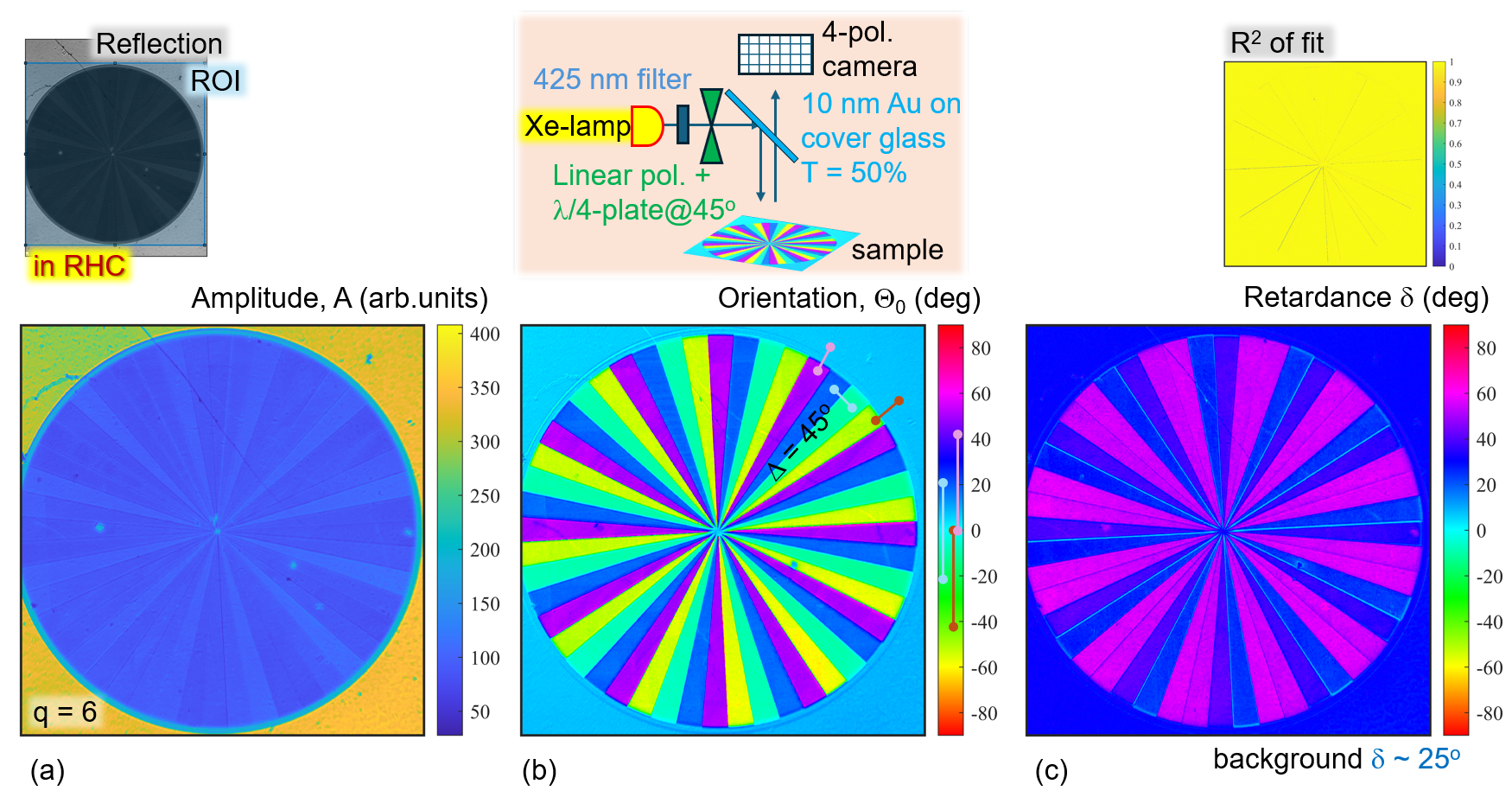}
\caption{\label{f-qu6R} Same sample, setup and focusing only in the Reflection mode; Xe-lamp illumination was used and the commercial dichroic mirror was changed to a 10~nm Au evaporated film on cover glass which had $T = 0.5$. Colour-markers in (b) show separation by $\Delta\theta_0= 45^\circ$. For reference, reflectance from flat region of sample from the same measurement was used. There was no back-reflecting mirror below the sample.
}
\end{figure*}

Next, both $T$ and $R$ modes of imaging with 4-pol. camera was carried out with the same sample placement, objective lens with higher $NA=0.5$ and in the illumination by RHC light using the very same polariser and filter set. The only difference was that a halogen lamp condenser was used for illumination in the $T$-mode and Xe-lamp in the $R$-mode (see insets in Figs.~\ref{f-qu6T} and \ref{f-qu6R}). Equation~\ref{e1} was applied for the both fits since it is derived for a circularly polarised light illuminating a generic optical retardance place (sample) and analysed with a linearly polarisers at the detector. Validity for the application Eq.~\ref{e1} for reflection is justified by the action of reflectance at the bottom of q-plate and diamond plate (see Fig.~\ref{f-sem}(a)). The $\pi$-shift is expected due to refractive index ratios and the E-fields are flipped to opposite orientation (on the same azimuth). Hence, the same refractive index is acting for the back-propagating light.

%____________________________fig 5
%____________________________fig 6
\begin{figure*}[tb]
\centering\includegraphics[width=1\textwidth]{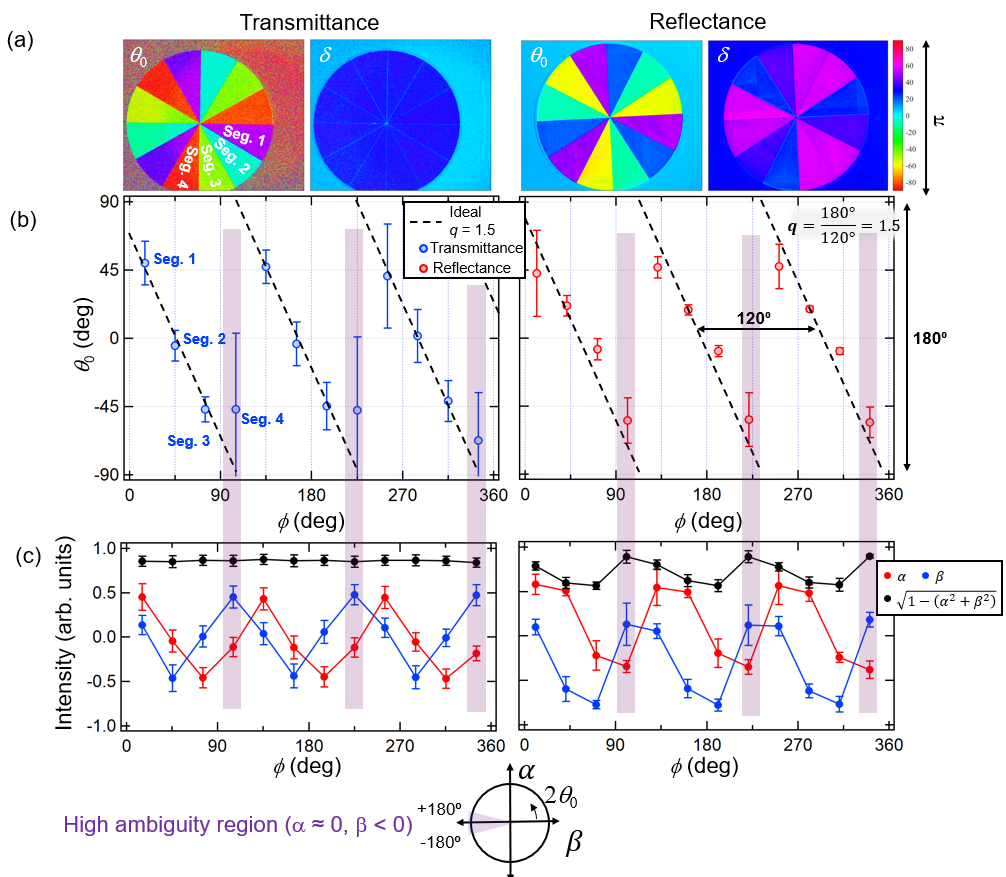}
\caption{\label{f-Stok} Segment-averaged value of $\theta_0$ for the $q=1.5$ plate: (a) Transmittance and Reflectance images calculated by Eqn.~\ref{e1}, (b) the $\theta_0$ vs. the azimuth angle $\phi$, and expressions of Stokes $S_0$ at selected angles (see text for expressions; averaging was carried out in polar coordinates as shown in Fig.~\ref{f-prot}). The vertical shadow-markers highlight the regions where $\theta_0$ is ambiguously defined since it is were the $\pi$-folding of theoretically expected to occur. The bottom marker shows the $(\alpha.\beta)$-plane. Note that $\alpha$ and $\beta$ is not directly correspond to Stokes parameter $S_1$ and $S_2$; see definition in the text.%\red{The vertical axis of (c) should be changed to "Intensity (a. u.)". The notation "High ambiguity ..." should be also changed to "High ambiguity region ($\alpha \approx 0$, $\beta < 0$)" . We can also remove notation "$S_1$-$S_2$ cross section of ..."} 
}
\end{figure*}

In transmission (Fig.~\ref{f-qu6T}), high fidelity determination of orientation in q-plate segments was obtained. The retardance value from the fit $\delta\simeq 30^\circ$ was obtained as in previous measurements with lower $NA = 0.13$. As expected, $\delta =0$ at the diamond plate locations (Fig.~\ref{f-qu6T}(c)). For fit analysis in refection (Fig.~\ref{f-qu6R}), the orientation readout was found not exactly corresponding to the actual structure with $\Delta\theta_0 = 45^\circ$ (Fig.~\ref{f-qu6R}(b)). The expected retardance in $R$, is expected to double from $T$ due to twice longer light propagation $2H$ in form birefringent region with the same birefringence $\Delta n$ (as discussed above). Indeed there were determined $\delta\simeq 60^\circ$, which is double of that in $T$-mode (Fig.~\ref{f-qu6T}). However, this was not consistent throughout all the q-plate segments. The background retardance $\delta\simeq\pi/8$ was present in $R$-mode, which is not expected and was absent in $T$-mode. The same $\pi/8$ retardance was in the q-plate segments, which were not following expected $\delta\sim 60^\circ$. From the amplitude $Amp$ (Fig.~\ref{f-qu6T}(a)), the q-plate has low reflectance as expected to the structure of $\sim\lambda/4$ (or retardance $45^\circ$ measured reliably in $T$-mode). Despite low-$R$, the fit was good with $R^2>0.9$ (inset in Fig.~\ref{f-qu6R}(c)). 

%The reason of orientation mismatch in $R$-mode from actual (confirmed in $T$-mode) %, could be is due to applicability of Eq.~\ref{e1}. It was derived for one handedness (helicity), but the handedness changed to opposite (right-to-left) in the reflection (since the direction of light propagation is inverted upon reflection, the spin angular momentum is conserved). Also, the 10~nm Au mirror on a cover glass is encoutered twice in the beam at 45$^\circ$ incidence affecting state of polarisation.}

By using Eqn.~\ref{e1}, one can determine the intensity at four different polariser angles of the 4-pol. camera $\phi=0,\pi/4,\pi/2,3\pi/4$: $S_0(\phi = 0) = 1 + \sin\delta\sin[2\theta_0]$, $S_0(\phi = \pi/4) = 1 - \sin\delta\cos[2\theta_0]$, $S_0(\phi = \pi/2) = 1 - \sin\delta\sin[2\theta_0]$, $S_0(\phi = 3\pi/4) = 1 + \sin\delta\cos[2\theta_0]$ (see plots in Fig.~\ref{f-prot}); here the 0.5 multiplier is omitted. Then $\alpha \equiv S_0(\phi=0)-S_0(\phi=\pi/2) = 2\sin\delta\sin[2\theta_0]$ and $\beta\equiv S_0(\phi=3\pi/4)-S_0(\phi=\pi/4) = 2\sin\delta\cos[2\theta_0]$ The local grating orientation azimuth (independent on retardance $\sin\delta$) is $\theta_0 = \frac{1}{2}\tan^{-1}\left(\frac{\alpha}{\beta}\right)$. Figure~\ref{f-Stok}(a) shows the images in $T$ and $R$-modes, which were processed to determine orientation of each segment $\theta_0$ from $\alpha/\beta$ as the azimuth $\phi$ angular position on the q-plate in (b). The $\alpha$ and $\beta$ are plotted in (c). All data are plotted as average and error bars corresponds to standard deviation $\pm\sigma$. Figure~\ref{f-Stok1} confirms robustness of the used approach on a more segmented $q=6$ sample. 

%Better understanding of orientation readout from reflection is required. Numerical modeling might help.
%\red{[...modeling to add ?..... to see difference we could setup fdtd: 1. plane-wave source is above sample ( 1-2 wavelengths) and monitor is 3 wavelengt above for the side and tip views 2) 10nm Au mirror reflect plane wave to the sample (1-2 wavelengths above)] and sice view as in case 1.}

\begin{figure*}[tb]
\centering\includegraphics[width=1\textwidth]{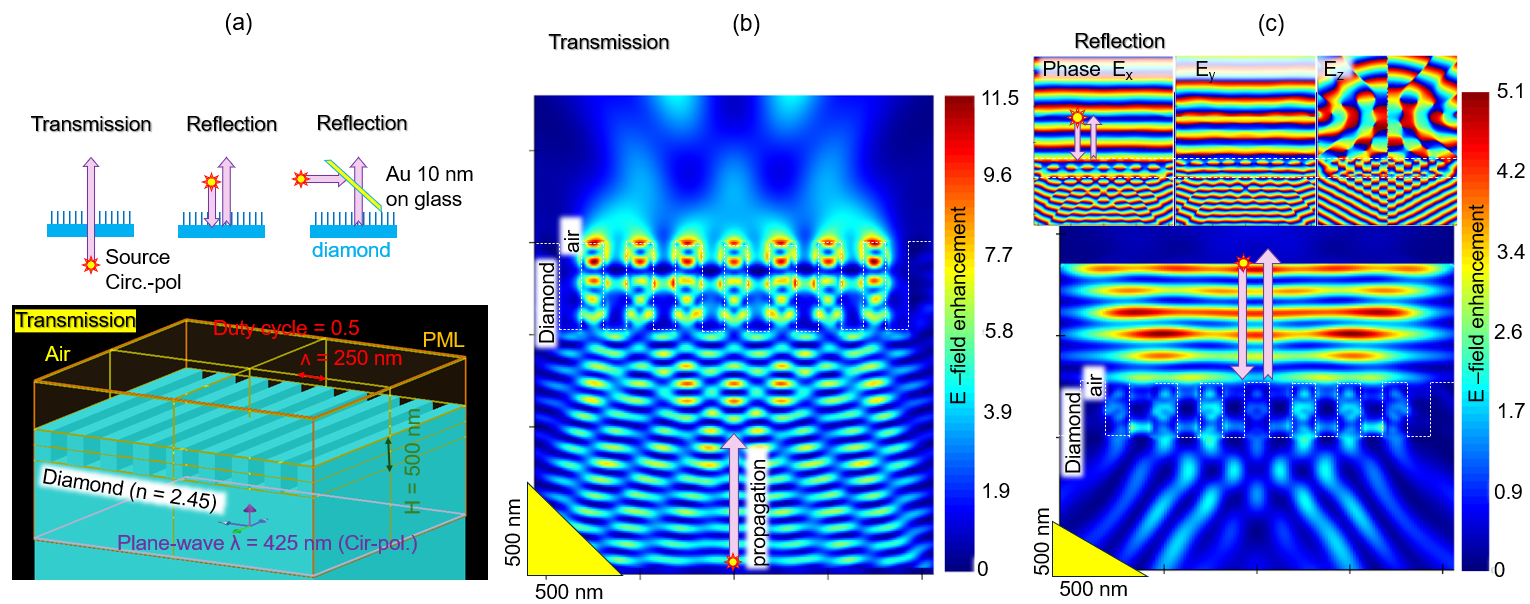}
\caption{\label{f-fdtd} FDTD calculations. (a) Geometries investigated for reflection of the circularly polarised transmission/reflection modes. The $T$-mode measurements recovered orientation of q-plates' segments. The $R$ mode without and with half mirror (10 nm Au on cover glass) for testing interference pattern (phase) changes. All calculations were carried out with perfectly matching layer (PML) to eliminate/reduce back-reflections into the calculation volume. (b) Transmission mode, side view intensity cross section (xz-plane); incident field $|E| = 1$. To reduce number of spurious reflections affecting the interference pattern, the light source was set inside diamond substrate. (c) Reflection with the source of circularly polarised light $\sim 5\lambda$ above the sample. The phase maps of $E_{x,y,z}$ components in the xz-plane; the color map blue-red is $-\pi... \pi$.    }
\end{figure*}

The finite difference time domain (FDTD) calculations were carried out to visualise light intensity distributions which reveals the interference, hence, phase changes for the $T$ and $R$ modes (Fig.~\ref{f-fdtd} with reference calculations for pure Au or \ce{SiO2} reflectors shown in Figs.~\ref{f-Au} and \ref{f-sio2}, respectively). For the $T$-mode, the orientation of q-plate segments was perfectly revealed. For $R$-mode, there was %a miss-determined 
the orientation~(Fig.~\ref{f-qu6R}) using Eq.~\ref{e1} was also correctly determined only with uncerntainy when the phase change is close to $\theta_0\simeq\pm\pi/2$. %The correct orientation was obtained by use of generic cosine-fit as described below. %\red{TBC.....}

Finally, the most generic fit to reveal absorption anisotropy (in transmission) by establishing orientation of the maximum absorbance $2\theta_0$ using the linearly polarised light $T(\vartheta)=Amp\times\cos[2\vartheta-2
\theta_0] + \mathrm{Offset}$ was applied for reflectance and circularly polarised illumination of the sample (Fig.~\ref{f-bingo}). This approach can be justified since the form birefringent grating structure is self-similar upon $\pi$ rotation, exactly as for the absorbers (see Fig.~\ref{f-fb}(b)). The fit was performed at the pixel level and the orientation $2\theta_0 \approx 45^\circ$ as expected for the actual sample. The large $\theta_0\simeq 112$ between two neighbouring segments yields to the same $2\theta_0 = 45^\circ$ due to $\pi$-folding, i.e., $2\theta_0 \approx 2\times 112^\circ - 180^\circ$ (Fig.~\ref{f-bingo}(b)). The fit has good fidelity $R^2>0.9$ even for the small amplitude $Amp$ and large offset. %was not appropriate for the circularly polarised light illumination in reflection nor due to the birefringent nature of the sample (very weak absorption). 

%____________________________fig 7
%____________________________fig 8
\begin{figure*}[tb]
\centering\includegraphics[width=1\textwidth]{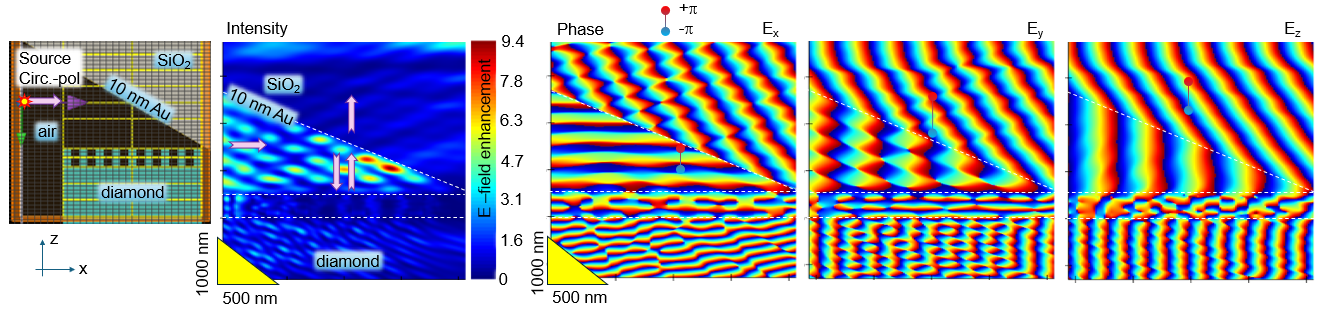}
\caption{\label{f-phase} FDTD calculations for the side-illumination by circularly polarised light. The phase maps for the $E_{x,y,z}$ fields; color heat map blue-red is $-\pi...\pi$. The refractive index of gold at 425~nm $n+i\kappa = 1.4532+i1.9512$.  }
\end{figure*}

\section{Discussion}\label{disco}

Another way to carry out analysis of the detected light after it passed a retarder (a birefringent sample) is via Stokes polarimetry~\cite{Otani}. The 4-pol. camera readout was incorporated into analytical expression~\cite{24ap2300471} as the following discussion. The Stokes vector after passing the $\lambda/4$-plate produce the right-hand circular (RHC) illumination onto the sample (retarder plate): 
\begin{equation}\label{e-asol}
S' = M_{ret}(\delta,\theta)\cdot S_{RHC} =
\begin{pmatrix}
        1 \\
        \sin{2\phi}\sin{\delta}\\
        -\cos{2\phi}\sin{\delta} \\
        \cos{\delta}
    \end{pmatrix}
\end{equation}
The experimentally detected intensity by 4-pol. camera is (polariser at $\phi$ angles):
\begin{equation}\label{eS}
S_0(\phi)\equiv I_\phi=\frac{1}{2}(S'_0+S'_1\cos2\phi + S'_2\sin2\phi).
\end{equation}
For the 4-pol. angles $\varphi$, the corresponding intensities reads: $I_0 = (S'_0 + S'_1)/2$, $I_{90} = (S'_0-S'_1)/2$, $I_{45} = (S'_0+S'_2)/2$, and $I_{-45} = (S'_0-S'_2)/2$. The Stokes parameters $S'$ are directly related to the measured intensity by the 4-pol. camera: $S'_0 =(I_{0} + I_{45} + I_{90} + I_{-45})/2$, $S'_1 = I_0 - I_{90}$, $S'_2 = I_{45} - I_{-45}$. There is an ambiguity with $S'_3$ parameter since two angles $\theta_{1,2}$ satisfy $\tan 2\theta = -S'_1/S'_2$: $\theta_1 = \arctan_2(-S'_1,S'_2)/2$ and $\theta_2 = \arctan_2(S'_1,-S'_2)/2$ (both can be calculated for comparison and cross check with common sense). The corresponding retardance $\delta_{1,2}$, which are expressed via Stokes parameters $S'_{1,2}$: $\delta_1 = \arcsin{(S'_1/\sin2\theta_1)} = \arcsin{(-\sqrt{S\mathrm{'}_{1}^2 + S\mathrm{'}_{2}^2})}$ and $\delta_2 = \arcsin{(S'_1/\sin2\theta_2)} = \arcsin{(\sqrt{S\mathrm{'}_{1}^2 + S\mathrm{'}_{2}^2})}$. Set of $(\delta_{1,2}, \theta_{1,2})$ satisfies $S'_3(\delta_1) = S'_3(\delta_2)$ and can be calculated using experimentally determined 4-pol. intensity images for $S'_{0,1,2}$. 

This shows that Stokes polarimetry is another method to determine the anisotropy using 4-pol. images. However the same challenge of change of handedness upon reflection is present and not explicitly incorporated in the derivation of Eq.~\ref{eS}. When circularly polarised light is used, one needs information for which depth $H$ it was probing birefringent structure and account for the same height $H$ in back-reflected light with changed handedness arriving at the 4-pol. camera.
Since Stokes analysis by Eqn.~\ref{eS} is made using image frames, rather numerical fit of each pixel, it is faster.

The demonstrated generic fit used for reflectance at the pixel level showed pattern recognition of form birefringent structures with modulo $\pi$ uncertainty (Fig.~\ref{f-bingo}). In remote sensing and imaging using reflection (e.g., Earth's observation), the $\pi$-folded absorbance or form-birefringence can be mistaken in the acquired image and other information is required for the distinction of the actual orientation. Yet, use of a circularly polarised illumination and 4-pol. capability in image acquisition are two key features required for a wider range of applications and real-time monitoring of order-disorder transitions and pattern formation, for example. 

%\red{Ideas: 1. to use eqn 5 for best fit and $Es \cos^2 + Ep \sin^2$. 2. can S be calculated from images algebra to flip RHC with LHC? probably not since S3 is not there, which is circ pol sensitive. 3. Null intensity setup with crossed circ-polariser and circ-analiser, then birefringence is visualised. }

\section{Conclusions and Outlook}

Simplification and faster processing of polarisation analysis in microscopy is demonstrated for detection/visualisation of structural anisotropy and alignment by using circularly polarised light produced by simple plastic polarizers placed on the sample's illumination port with a spectral band-pass filter for selecting the wavelength of interest. Complex form birefringent structures of q-plates made on diamond membrane were imaged in transmission and showed expected alignment with the actual pattern. In reflection, a circular polarisation undergo complex changes upon reflection from the half-mirror, then back-reflection from sample, and passing the same half-mirror before 4-pol. camera image detection. %This affects recognition of the 
However, the structure's azimuth can be recovered from reflection using analytical expression based on Stokes parameter $S_0$ (same as for transmission). %However, by using 
Also, a simple $\cos(2\theta)$-fit can be used for the determination of structure orientation %can be determined 
in reflection with $\pi$-folding uncertainty.

%Future 
The demonstrated simplification of structural analysis in reflection mode upon illumination of sample with circularly polarised light based on Stokes polarimetry principles opens applications at wide range of EM-spectrum. The image processing could be %experimentally explored 
used for faster data analysis instead of fitting which has to be carried out at each pixel. The reflection based anisotropy and orientation analysis is promising %for other spectral ranges and can be used 
for the Earth observation where instantaneous detection of four polarisation images in reflection is key requirement (upon illumination with circularly polarised light).

\small\begin{acknowledgments}
S.J. acknowledges support via ARC DP240103231 grant and Nanophorb (2018-2022) - Nanofabrication technologies towards advanced control of the photon angular momentum - from the National Center for Scientific Research (Le Centre National de la Recherche Scientifique), France.  We are grateful  to Etienne Brasselet for design of q-plates. H-H.H. and S.J. are grateful for a research stay at the Laser Research Center, Vilnius University, in 2025. Preliminary works leading to this project were performed using synchrotron-based FTIR technique on the Infrared Microspectroscopy (IRM) beamline at the Australian Synchrotron, part of ANSTO, through merit-based beamtime proposals (ID. 20039, 20635, and 22505). 
\end{acknowledgments}

%\nocite{*}
%\newpage
\bibliography{aipsamp}% Produces the bibliography via BibTeX.

\appendix
\setcounter{figure}{0}\setcounter{equation}{0}
\setcounter{section}{0}\setcounter{equation}{0}
\makeatletter 
\renewcommand{\thefigure}{A\arabic{figure}}
\renewcommand{\theequation}{A\arabic{equation}}
\renewcommand{\thesection}{A\arabic{section}}

\section{Imaging in $T$- and $R$-modes}

\begin{figure*}[h!]
\centering\includegraphics[width=1\textwidth]{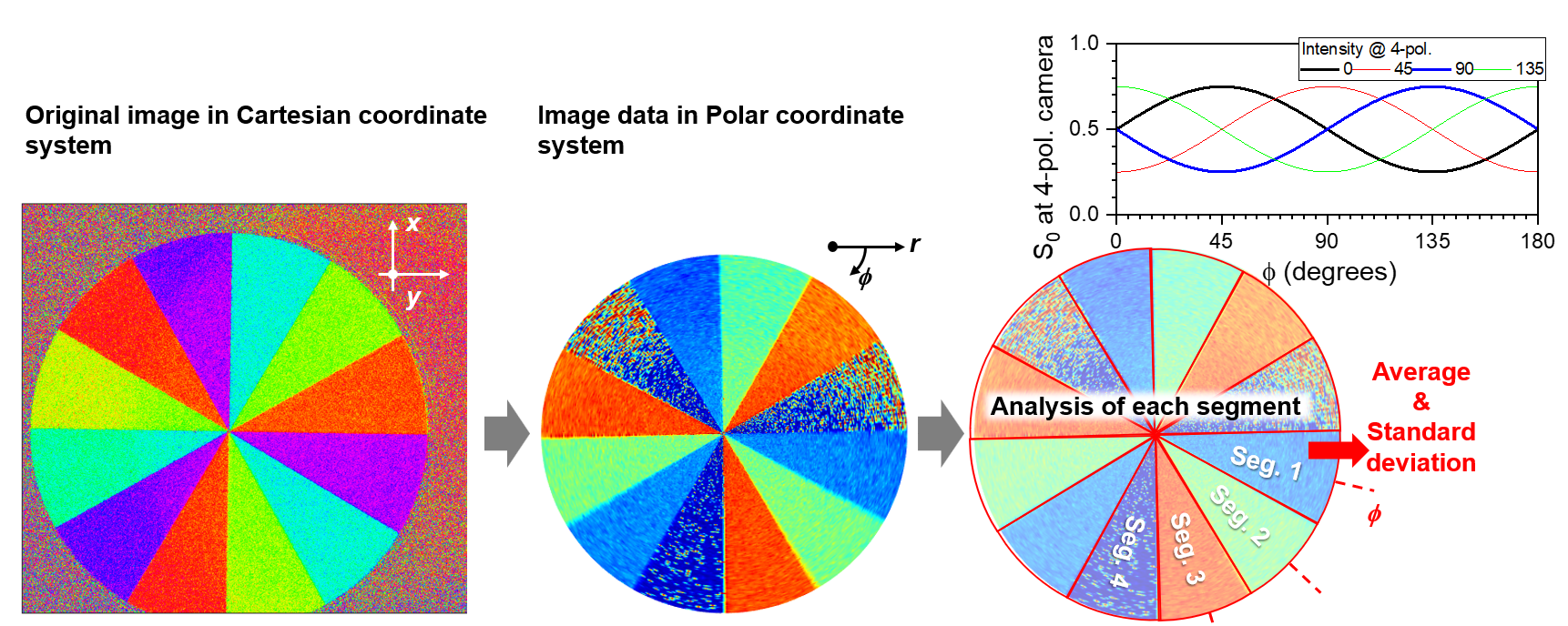}
\caption{\label{f-prot} Exemplary image data of $\theta_0$ for the analysis. Change of Cartesian $(x,y)$ to Polar $(r,\phi)$ coordinate system. Averaging and image analysis were carried out using Stokes $S$ parameters, rather the fit. The inset-plot shows intensity on each of 4-pol. camera segments when the birefringence $\sin\delta = 0.5$ (Eqn.~\ref{e1}).
}
\end{figure*}

Figure~\ref{f-prot} illustrates the applied protocol to calculate the average and standard deviation of the orientation angle $\theta_0$ on selected segment in both $T$ and $R$-modes. This was achieved by working in polar coordinates which allowed a simple math. The plot inset shows how the 4-pol. camera segments project the incident intensity/power based on Eqn.~\ref{e1}; the $\phi$ is the polariser orientation angle. 

\begin{figure*}[h!]
\centering\includegraphics[width=1\textwidth]{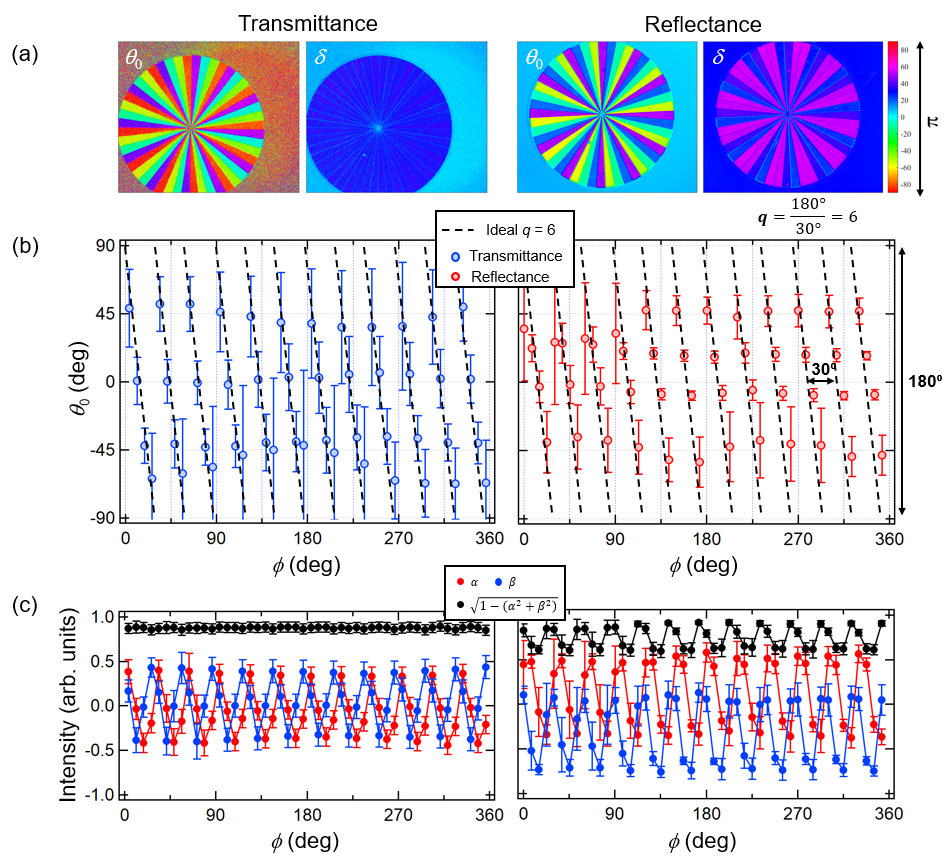}
\caption{\label{f-Stok1} Segment-averaged value of $\theta_0$ for $q=6$ plate: (a) Transmittance and Reflectance images calculated by Eqn.~\ref{e1}, (b) the $\theta_0$ vs. the azimuth angle $\phi$, and (c) expressions of Stokes $\alpha,\beta$ expressed via $S_0$ at selected angles. %\red{The vertical axis of (c) should be changed to "Intensity (a. u.)".}
}
\end{figure*}

Figure~\ref{f-Stok1} shows the applied analysis for the $T$ and $R$ images for $q = 6$ plate, where nano-gratings have different orientation every 30$^\circ$. The $\pi$-folding of $\theta_0$ creates regions of angles around $\pm\pi/2$ where the largest uncertainty exist for determined orientation angle $\theta_0$. Importantly, for the averaging the $\sin\delta$ was taken from experimental fit by Eqn.~\ref{e1} (Fig.~\ref{f-Stok1}(a)). This is reflected in the amplitude shift in $\alpha$ and $\beta$ in Fig.~\ref{f-Stok1}(c). %\red{[Note: since RCP-circular becomes elliptical, the $\alpha$-red is shifted in respect $\beta$-blue.]}

 %____________________________fig a1
 %____________________________fig a2

 %____________________________fig a3
\begin{figure*}[h!]
\centering\includegraphics[width=1\textwidth]{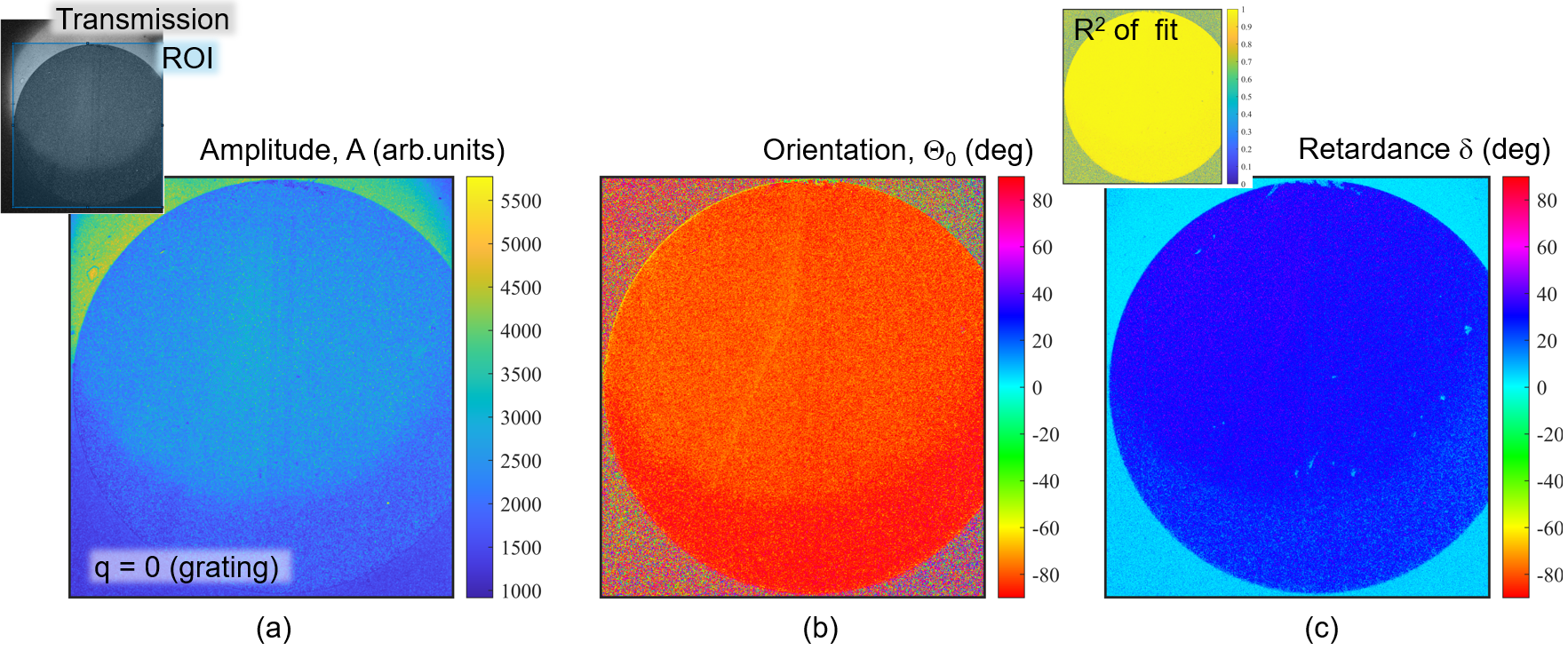}
\caption{\label{f-q0} Transmittance $T$ (RHC illumination) of $q=0$ plate (grating) imaged by 4-pol. camera at numerical aperture $20^\times NA = 0.5$ and fitted by $Fit(\phi)=Amp[(1-\sin\delta\sin{[2(\phi - \theta_0)]})/2]$ (see panels (a-c) for the fit parameters amplitude, orientation (highest $T$), and retardance, respectively: $Amp, \theta_0, \delta$ calculated for each pixel).
}
\end{figure*}
 %____________________________fig a4
\begin{figure*}[h!]
\centering\includegraphics[width=1\textwidth]{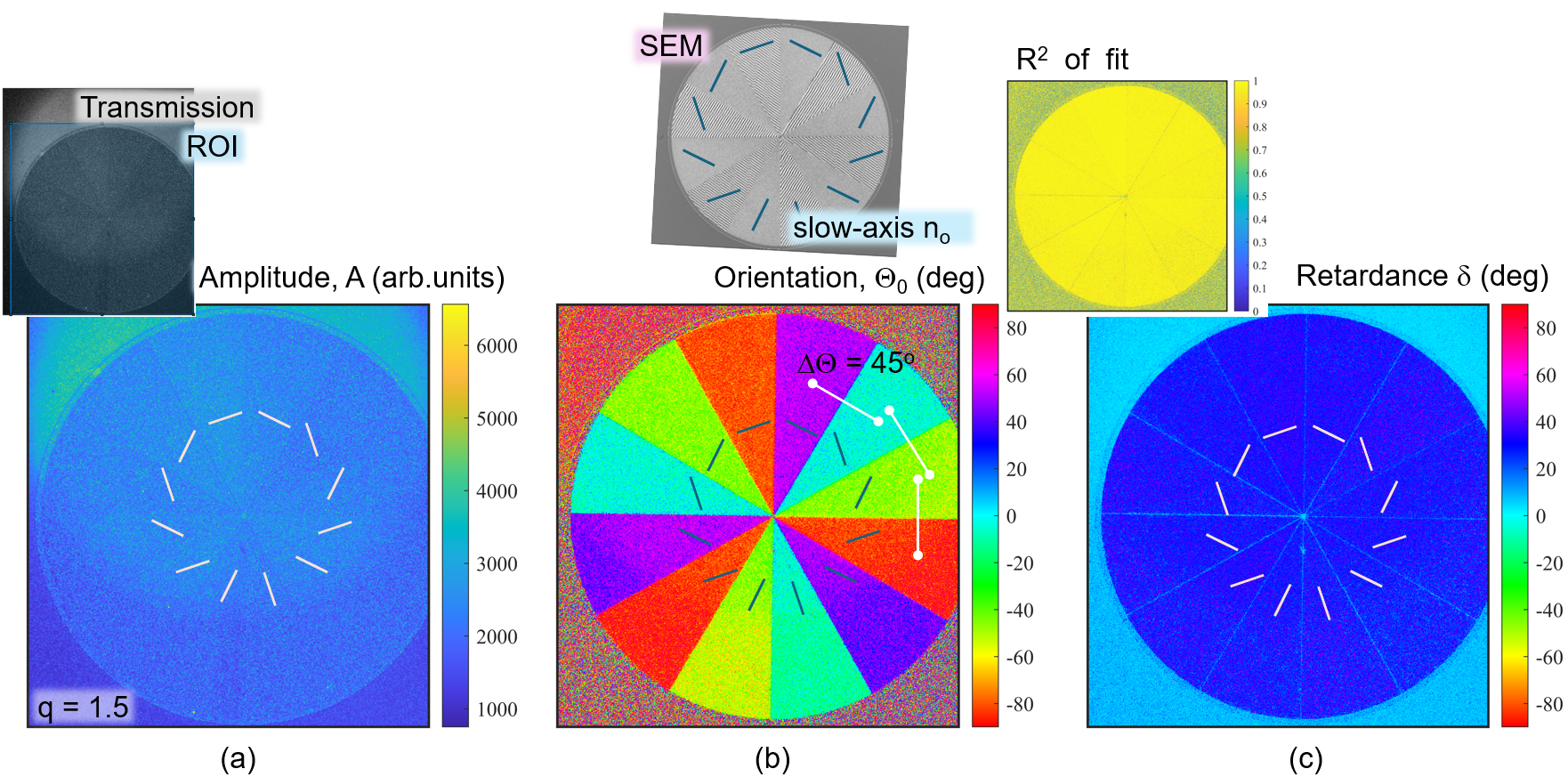}
\caption{\label{f-qT} Transmittance $T$ (RHC illumination) of $q=1.5$ plate imaged by 4-pol. camera at numerical aperture $20^\times NA = 0.5$ and fitted by $Fit(\phi)=Amp[(1-\sin\delta\sin{[2(\phi - \theta_0)]})/2]$ (see panels (a-c) for the fit parameters amplitude, orientation (highest $T$), and retardance, respectively: $Amp, \theta_0, \delta$ calculated for each pixel). Top-inset in (b) shows the SEM image, which is aligned with optical images and shows the actual orientation of slow axis ($n_o$).}
\end{figure*}
 %____________________________fig a5
\begin{figure*}[h!]
\centering\includegraphics[width=1\textwidth]{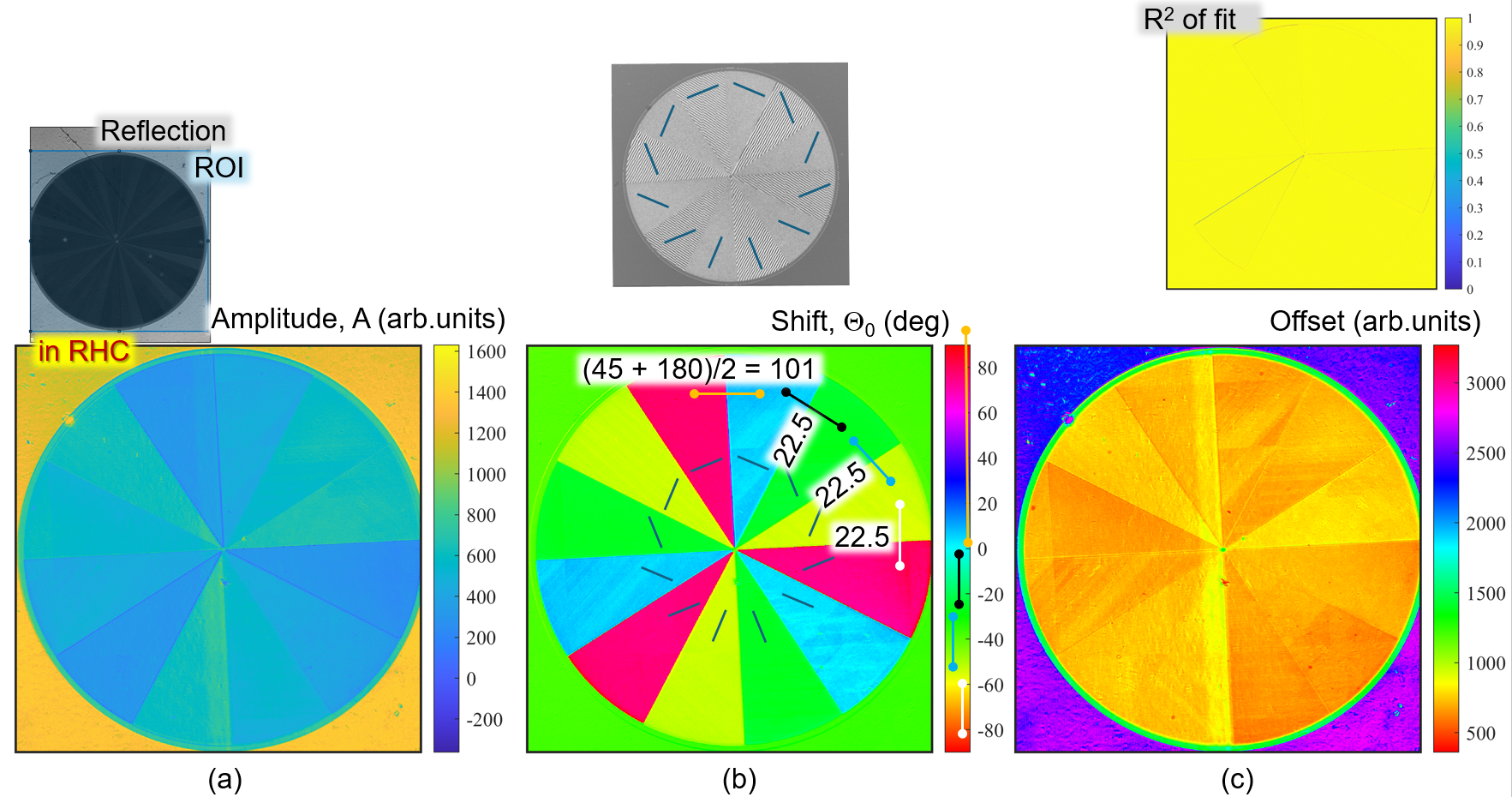}
\caption{\label{f-bingo} Reflectance $R$ (in RHC illumination) of $q=1.5$ plate imaged by 4-pol. camera at numerical aperture $20^\times NA = 0.5$ and fitted by $R(\phi)=Amp\times\cos(2\vartheta -2\theta_0)+Offset$ (see panels (a-c) for the fit parameters amplitude, orientation ($2\theta_0$ at the highest $T$), and retardance, respectively: $Amp, \theta_0, Offset$ calculated for each pixel). Top-inset in (b) shows the SEM image, which is aligned with optical images and shows the actual orientation of slow axis ($n_o$).}
\end{figure*}
%____________________________fig a6
\begin{figure*}[tb]
\centering\includegraphics[width=1\textwidth]{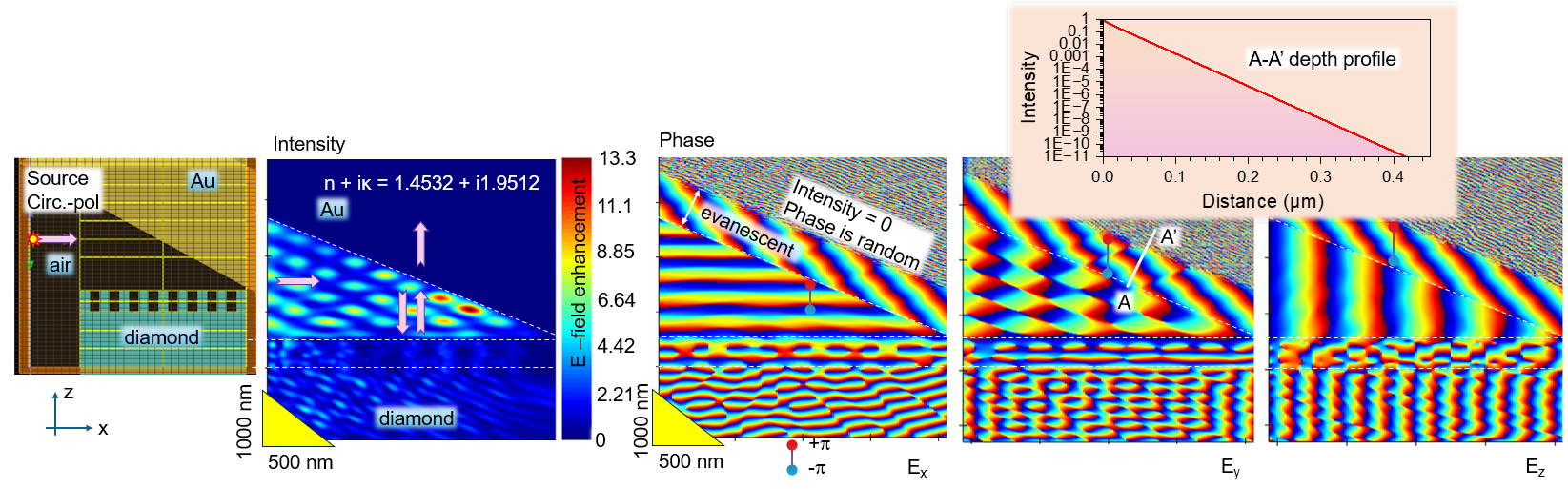}
\caption{\label{f-Au} FDTD calculations for the side-illumination by circularly polarised light with all reflector from Au (10 nm Au on Au block). The phase maps for the $E_{x,y,z}$ fields; color heat map blue-red is $-\pi...\pi$. The refractive index of gold at 425~nm $n+i\kappa = 1.4532+i1.9512$ and reflection phases $\vartheta_p = 38.43^\circ$ and 
$\vartheta_s = -141.57^\circ$ for s-/p-polarisations. The region of random phase inside sub-surface of Au is where intensity is close to zero; see top-inset a cross-sectional intensity profile perpendicular to the surface. }
\end{figure*}
%____________________________fig a7
\begin{figure*}[tb]
\centering\includegraphics[width=1\textwidth]{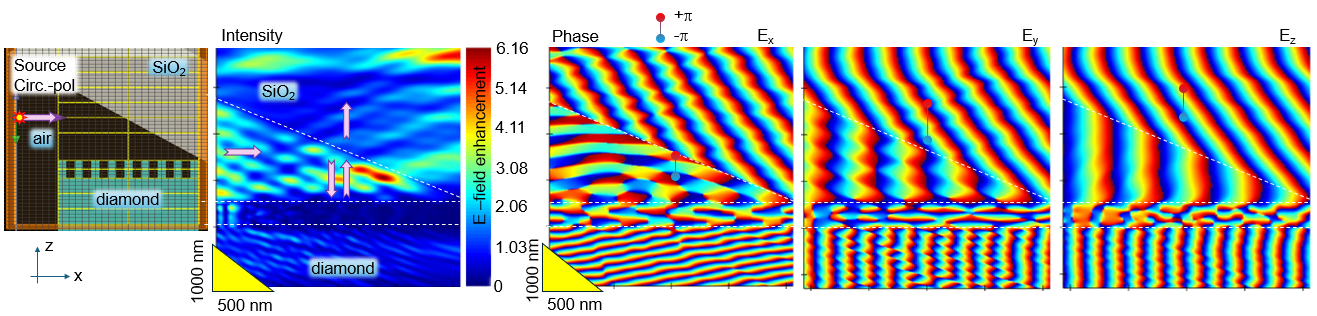}
\caption{\label{f-sio2} FDTD calculations for the side-illumination by circularly polarised light with all reflector from \ce{SiO2} (10 nm \ce{SiO2} on \ce{SiO2} block). The phase maps for the $E_{x,y,z}$ fields; color heat map blue-red is $-\pi...\pi$. The refractive index of silica at 425~nm $n = 1.45$. }
\end{figure*}

Imaging at larger $NA = 0.5$ is shown for the grating ($q = 0$ no azimuthal dependence of birefringence) in transmission (Fig.~\ref{f-q0}) and for $q = 1.5$ in Fig.~\ref{f-qu6T}. Both show high fidelity fits by Eq.~\ref{e1} and determined parameters of $Amp, \theta_0, \delta$. For reflection image, a generic expression of $\pi$-folded anisotropy (absorption of birefringence) was used for the fit by $R(\vartheta) = Amp\times\cos(2\vartheta - 2\theta_0)$ with extracted parameters in Fig.~\ref{f-qu6R}; single-angle equivalent expression is $R(\vartheta) = 2Amp\times\cos^2(\vartheta - \theta_0) - Amp$ (since $\cos2\theta = 2\cos^2\theta - 1$). In this presentation the orientation azimuth of $2\theta_0\simeq 45^\circ$ was obtained. Interestingly, for the one orientation where $\pi$-folding had to be applied, the amplitude $Amp$ was the smallest. In the fitting procedure there was no restriction on the sign of parameters, i.e., negative $Amp$ was allowed for the fit. As it was observed in other measurements, the negative $Amp$ was associated with phase change by $\pi$, which was the case for this $q = 1.5$ sample. 

\section{Link between ellipsometry and 4-pol. in $R$-mode}

The generic ellipsometry formula $\rho=\tan\Psi e^{i\Delta}$ can be expressed via Euler's equation for the reflection (for field) coefficients: $\frac{r_p}{r_s} = \frac{|r_p|}{|r_s|}\cos\Delta + i\frac{|r_p|}{|r_s|}\sin\Delta$. The magnitude $\tan\Psi \equiv \frac{|r_p|}{|r_s|}$ is $\left(\frac{|r_p|}{|r_s|}\right)^2 = \frac{|r_p|}{|r_s|}\cos^2(\delta_p-\delta_s) + \frac{|r_p|}{|r_s|}\sin^2(\delta_p-\delta_s)$. This reflectivity expression could be used for non-normal incidence onto the sample. Similar to the ellipsometry reflectance expression above, the reflected light in attenuated total reflection (ATR) mode of polarisation analysis showed that generic equation of $T_{ATR} = A_p\cos^2(\vartheta-\vartheta_p) + A_s\sin^2(\vartheta-\vartheta_s)$ describes the polarisation dependence of reflected intensity~\cite{21as7632}. The 4-pol. method can be applied for analysis of reflected light where the amplitudes of s-/p-pol. are changed and accounted for by $A_{s,p}$ as well as their phases $\vartheta_{s,p}$. It is noteworthy to mention the influence of objective lens to the possible phase change upon reflection. For example, in the case of the Cassegrain reflective objective lens used in reflection, the illumination of the sample takes place at a considerable large ($\alpha_i\sim 20-40^\circ$) angle of incidence when $NA\geq 0.5$ is used ($NA = n\sin\alpha_i$, with $n = 1$ in air), while the reflected part of light collected with the same objective lens on the other half of Cassegrain. In such a case, there is a phase shift $\sim 30^\circ$ induced upon reflectance from the reference gold mirror used as a reference to normalise intensities~\cite{21n3247}. This has to be considered when reference is used for normalisation of amplitudes, however, it affects the phase as well. Convention of angle-$0^\circ$ can be made using transmitted light through the same objective lens (without the sample) for the orientation/phase reference.  

\section{Finite-Difference Time Domain simulations}

Figure~\ref{f-fdtd} shows the FDTD simulations for the actually used 10~nm Au mirror reflector. For reference to compare the phase changes, reflections from pure Au and \ce{SiO2} for the exactly same geometry are shown in Figs.~\ref{f-Au} and \ref{f-sio2}, respectively. Interestingly, refractive index of Au at 425~nm is 1.4532 is similar to that of \ce{SiO2} 1.45, hence, the evanescent phase in sub-surface region of Au mirror has a similar phase structure as inside pure silica reflector. 

\end{document}